\documentclass[12pt]{article}

\usepackage{graphicx}
\usepackage{epstopdf}

\def \lsim{\mathrel{\vcenter
     {\hbox{$<$}\nointerlineskip\hbox{$\sim$}}}}

\newcommand{\beq}{\begin{equation}}
\newcommand{\eeq}{\end{equation}}
\newcommand{\beqa}{\begin{eqnarray}}
\newcommand{\eeqa}{\end{eqnarray}}
\newcommand{\beqar}{\begin{eqnarray*}}
\newcommand{\eeqar}{\end{eqnarray*}}

\begin{document}
\thispagestyle{empty}

\hfill{\sc UG-FT-250/09}

\vspace*{-2mm}
\hfill{\sc CAFPE-120/09}

\vspace{32pt}
\begin{center}

\textbf{\Large Cosmic-ray knee and flux of secondaries from\\
interactions of cosmic rays with dark matter} 
\vspace{40pt}

Manuel Masip$^{1}$, Iacopo Mastromatteo$^{1,2}$
\vspace{12pt}

\textit{
$^{1}$CAFPE and Departamento de F{\'\i}sica Te\'orica y del
Cosmos}\\ 
\textit{Universidad de Granada, E-18071 Granada, Spain}\\
\vspace{8pt}
\textit{$^{2}$International School for Advanced Studies (SISSA)\\ 
Via Beirut 2-4, I-34014 Trieste, Italy\\
}
\vspace{16pt}
\texttt{masip@ugr.es, iacopomas@infis.univ.trieste.it}
\end{center}

\vspace{40pt}

\begin{abstract}

We discuss possible implications of a large interaction
cross section between cosmic rays and dark matter particles
due to new physics at the TeV scale. In particular, 
in models with extra dimensions and a low fundamental 
scale of gravity the cross section grows very 
fast at {\it transplanckian} energies. 
We argue that the knee observed in the cosmic ray flux could
be caused by such interactions. We show that this hypothesis
implies a well defined flux of secondary gamma rays that
seems consistent with MILAGRO observations.

\end{abstract}

\newpage

\section{New physics at the TeV scale}

We know from collider experiments that there are three basic 
interactions between elementary particles. These interactions
are understood in terms of a $SU(3)_C\times SU(2)_L\times U(1)_Y$
gauge symmetry, and have been confirmed
by all data during the past decades: the standard
model is basically {\it correct} up to  energies around  
$200$ GeV. On the other hand, we also observe 
gravitational interactions. Their strength is set by Newton's
constant, which in natural units defines the Planck mass, 
$M_P=G_N^{-1/2}$. Gravity is much weaker than gauge interactions
and not detectable at colliders. It has been tested only at
macroscopic distances, in processes involving the exchange
of {\it quanta} of up to $10^{-13}$ GeV. 

What do we expect
at higher energies? If we extrapolate what we {\it know} in a 
straightforward way, we find that the three gauge couplings
have log corrections that point towards a grand unification
scale at $M_X\approx 10^{16}$ GeV. Gravity is different,
it grows quadratically with the energy and becomes of order one
at the Planck scale, $M_P\approx 10^{19}$ GeV. Below $M_P$ 
one needs a consistent framework for the four
interactions, and string theory is the only available candidate.
The LHC is going to explore energies of up to 1 TeV. It could 
find, for example, supersymmetry, a discovery for the next
decades that would provide consistency to the whole picture. 
But such discovery would leave us still very far from the fundamental
scale. String theory and quantum gravity are in this framework
{\it non-reachable}, almost non-physical.

However, this is not the only possibility. In a different framework
that has been discussed a lot recently the fundamental scale of
gravity ($M_D$) is pushed down to the TeV. This can be done, for example,
with flat extra dimensions that accelerate
the running of $G_N$, or with a warped metric, the popular 
Randall-Sumdrum models\cite{ArkaniHamed:1998rs}. 
In any case, within this framework the 
LHC could see exciting physics, maybe even a hint of the string
scale itself. The {\it transplanckian} regime ($s\gg M_D^2$)
would probably
be not accessible there, but it would be clearly at the reach
of very energetic cosmic rays. Collisions in this regime are really
different from what we have seen so far in colliders. In particular,
the spin 2 of the graviton implies that gravity becomes strong and
dominates over gauge interactions at distances that increase with
$\sqrt{s}$.

The range of energies that cosmic rays provide is very wide, 
exceeding 
the 14 TeV to be reached at the LHC. In the
collision of a cosmic proton with a dark matter particle $\chi$ in
our galactic halo we have
\beq
\sqrt{s}=\sqrt{2m_\chi E}\lsim 10^7\;{\rm GeV}\;.
\eeq
One would obtain even higher energies, up to $10^{11}$ GeV, in 
the head on collision of two cosmic rays\cite{Draggiotis:2008jz}: 
these are the
most energetic elementary processes that we know are
occurring in nature, 
and would be clearly transplanckian within the TeV gravity 
picture. Here we will focus on the first type of processes.

\section{Transplanckian collisions}

Can one calculate a cross section at $\sqrt{s}\gg M_D$ without
knowing the details about the fundamental theory? The answer
is {\it yes} as far as the fundamental
theory does not change the long distance properties of 
gravity at these transplanckian energies. It
is the case, for example, in string theory, where the Regge 
behaviour implies that at $s\gg M_s^2$ only the low $t$
(forward) contributions of the massless string modes survive. 
And due to the spin 2 of the graviton, in this regime gauge 
contributions are negligible, only gravity matters.

In a collision at transplanckian energies we expect two
basic processes\cite{Illana:2005pu}. At small impact parameters we expect 
{\it capture}, the collapse of the two particles into a
mini black hole of mass $M\approx \sqrt{s}$ and radius
\beq
R\approx 
\left({M\over M_D}\right)^{1\over n+1}
     {1\over M_D}\;.
\label{rh}
\eeq
At larger impact parameters we expect processes where the
incident particle transfers a small fraction $y$ of its energy  
and keeps going. These elastic processes
can be calculated in the eikonal approximation, that provides
a resummation of ladder and cross-ladder contributions.

An important observation is that these are long-distance
processes, the typical distance is   
larger than $1/M_D$ and grows with the energy. To {\it see}
quantum gravity, string theory or even a $Z$ boson the incident
particle needs to go to short distances (of order $1/M_{Z,S,D}$)
{\it inside} the black hole horizon, so all these details 
become irrelevant. The higher the energy in the collision, 
the more reliable is the estimate based on classical gravity 
(strongly coupled but tree level).

Another important point is that, although the typical distance
is longer than $1/M_D$, it is still shorter than the proton 
radius and the exchanged gravitons {\it see} the partons inside
the proton. A parton carrying
a fraction $x$ of the proton momentum hits $\chi$ and, as a result,
the proton breaks into a scattering parton or a black hole
plus the proton remnant. From the analysis of these jets using
HERWIG we obtain

{\it (i)} The scattering parton and the proton remnant define
jets giving a very similar spectrum of stable particles.
This spectrum is only mildly sensitive to 
the fact that the parton may be a quark or a gluon.

{\it (ii)} In the center of mass frame of the two jets 
the final spectrum of stable particles 
is dominated by 
energies around 1 GeV, almost independently of the energy
of the parton starting the shower. 

{\it (iii)} The stable species (particle plus antiparticles) 
are produced with a frequency $f_i$ that is mostly independent
of the energy or the nature of the two jets. We obtain an
approximate 
55\% of neutrinos, a 20\% of photons, a 20\% of electrons, and
a 5\% of protons.

{\it (iv)} The spectrum of stable particles resulting from
a mini BH in its rest frame is very similar to the one obtained 
from the quark and gluon jets. 

A parametrization of the final spectra of stable particles from
quark and gluon jets and from black hole evaporation can be
found in\cite{Masip:2008mk}.

\section{Secondaries from collisions of cosmic rays with dark matter }

Let us now discuss if there is any observable effects from 
these processes. When a ultrahigh energy cosmic ray reaches the
Earth coming from
outside the galaxy, it has crossed a certain dark matter column density 
$x$. The probability of interaction is just
\beq
p(x)\approx {\sigma\; x\over m_\chi} \;.
\label{prob}
\eeq
Since the depth $x$ from the border
of the galaxy can vary in a factor of ten, more
cosmic rays will interact from deeper directions, which could
imply an anisotropy in the flux of extragalactic cosmic ray that has
not been observed. We find, however, that for cross sections up to
the mbarn and for the expected 
dark matter densities the probability of interaction
is too small to produce  an observable effect.

The effect on lower energy cosmic rays, however, could be more
relevant. The crucial difference is that cosmic rays of
energy below $10^8$ GeV are trapped inside the galaxy by
random magnetic fields of order $\mu$G. Their trajectory from
the source to the Earth is not a straight line, it is more
similar to a random walk. The depth that they face grows with time,
and a fraction of them could interact with a dark matter particle
before reaching the Earth. Now, in these models the cross section
grows very fast at center of mass energies above $M_D$, so there
could be a critical energy giving a cross section large 
enough for cosmic rays to interact. At larger  energies the 
interaction would break them and produce an effect that could 
explain the {\it knee} (the change in the spectral index from 
$-2.7$ to $-3$) observed in the cosmic ray spectrum.

If gravitational interactions were responsible for
this change, then there 
would be a flux of secondary particles that could be
readily estimated. Let us assume that, on absence of
gravitational interactions, the flux 
$\propto E^{-2.7}$ would have extended 
up to $10^{8}$ GeV. This means that the flux 
\beq
\Phi_{N}\approx \int_{10^6{\rm\; GeV}}^{10^8{\rm\; GeV}}
{\rm d}E\; 1.8 \; (E^{-2.7}-10^{1.8} E^{-3})
\; {\rm {nucleons\over cm^2\;s\;sr}}
\eeq
had been {\it processed} by these interacions into 
secondary particles of less energy. In Fig.~1 we plot 
the fluxes of secondary protons and gamma rays together 
with the flux of dark matter particles boosted by 
eikonal scatterings.
\begin{figure}
\begin{center}
\includegraphics[width=0.5\linewidth]{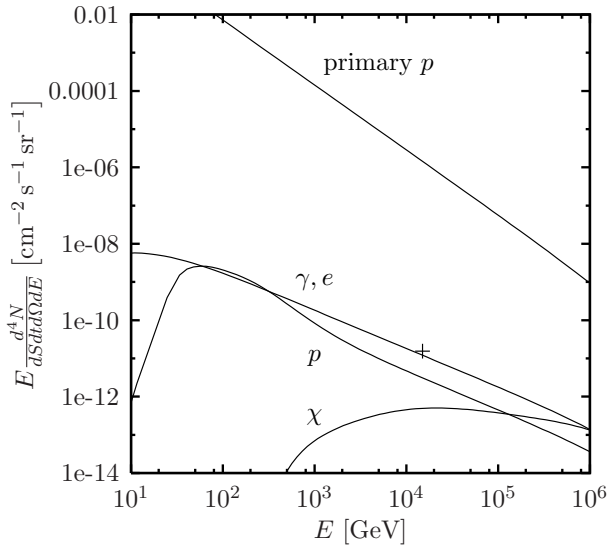}
\end{center}
\caption{Secondary fluxes from $p$--$\chi$ gravitational collisions
 for $n=6$, $M_D=5$ TeV and $m_\chi=200$ GeV. The point at 15 TeV
 indicates the gamma-ray flux measured by MILAGRO.
\label{fig7}}
\end{figure}
The flux of $e=e^++e^-$ is similar to the 
photon flux, although
the propagation effects (synchroton emission, etc.) that may
distort the spectrum have not been included. 
Recent data from PAMELA\cite{Adriani:2008zr} 
signals an excess in the positron flux 
above 10 GeV, although the contribution that we
find seems well below these data. 
We add in the plot the diffuse gamma-ray flux 
measured by MILAGRO\cite{Abdo:2008if} at energies around 15 TeV, 
which seems to indicate an excess versus the expected values from 
some regions in the galactic plane. 
The contribution
that we find could explain anomalies in the gamma-ray flux
above 10 GeV or in the positron and antiproton fluxes 
above 1 TeV. The diffuse photon flux that we obtain
is always around MILAGRO data and 
proportional to $E^{-2}$ at energies between 100 and $10^6$ GeV
for any values of the dark matter mass and the 
number of extra dimensions. 

\section{Summary}

Strong gravity at the TeV scale would affect the 
propagation of the most energetic cosmic rays. In particular,
cosmic protons could interact with the WIMP $\chi$ that 
constitutes the dark matter of our universe.
These interactions could break the incident proton and produce a
deflection in the flux (the cosmic ray knee), together
with a flux of secondary antiparticles and gamma rays.
The analysis of the cross sections and dark matter 
densities required for this hypothesis to work will
be presented elsewhere\cite{Barcelo:2009uy}. In any case, 
it is puzzling that the change in the 
spectral index in the flux appears at center of
mass energies $\sqrt{2m_\chi E_{knee}}\approx 10$ TeV,
where the new physics is expected. 

\section*{Acknowledgments}
This work has been supported by MEC of Spain 
(FPA2006-05294) and by
Junta de Andaluc\'\i a (FQM-101 and FQM-437).
IM acknowledges a fellowship from SISSA (2007JHLPEZ).

\end{document}